\pdfoutput=1
\documentclass[11pt,a4paper]{article}

\usepackage[utf8]{inputenc}
\usepackage[T1]{fontenc}
\usepackage{lmodern}
\usepackage{amsmath,amssymb}
\usepackage{amsthm}
\usepackage{graphicx}
\usepackage[hidelinks]{hyperref}
\usepackage{url}
\usepackage{booktabs}
\usepackage{natbib}
\usepackage[margin=1in]{geometry}
\usepackage{enumitem}
\usepackage{xcolor}

\newtheorem{proposition}{Proposition}

\title{From Prediction to Intervention: The Evolution of AI in Biomedicine}

\author{
  Andrew Feinberg,
  Aleksandr Sarachakov,
  Viktor Svekolkin,
  Alexander Bagaev,\\
  Ferran Prat,
  Michael Feinberg\textsuperscript{*}
  \\[6pt]
  \normalsize BostonGene Corporation, Waltham, MA, USA \\[4pt]
  \small\textsuperscript{*}Corresponding author: \texttt{michael.feinberg@bostongene.com}
}

\begin{document}

\maketitle

\begin{abstract}
Artificial intelligence has advanced rapidly in biomedicine through large-scale multimodal data integration, enabling increasingly accurate prediction of clinical outcomes and patient stratification. These systems, however, remain fundamentally observational: they learn statistical associations from historical data and operate within previously observed biological and clinical states, limiting their ability to generalize to novel therapies or unobserved interventions.

We argue that AI in biomedicine is undergoing a structural transition. As biomedical decision-making increasingly depends on reasoning about intervention rather than extrapolation from past observations, predictive architectures become structurally insufficient. Systems that learn from historical data cannot, by construction, represent how biological systems evolve under perturbation, and therefore cannot reliably support decision-making in the presence of novel interventions.

We introduce a conceptual framework distinguishing observational and interventional intelligence and define disease-level models as systems that explicitly represent the state, dynamics, and intervention response of biological processes. These models enable a shift from inference to simulation---reasoning about what will happen under intervention rather than what is likely based on the past.

This transition also implies a shift in where value is created: from data processing and prediction toward systems that support and define decision-making under intervention. It follows directly from the structure of biomedical decision-making and defines the next stage of AI in medicine. Systems that cannot model intervention will be structurally excluded from decision-making.
\end{abstract}

\medskip
\noindent\textbf{Keywords:} biomedical AI; observational intelligence; interventional intelligence; disease-level models; simulation; drug development; clinical decision support.

\section{Introduction: When Prediction Is Not Enough}
\label{sec:introduction}

Artificial intelligence has achieved significant success in biomedicine, particularly in oncology, where the integration of genomic, transcriptomic, imaging, and clinical data has enabled increasingly accurate prediction of patient outcomes, disease progression, and therapeutic response. Large-scale multimodal models now support tasks ranging from patient stratification to clinical trial optimization, and in many settings outperform traditional statistical approaches\citep{esteva2019guide,topol2019high,moor2023foundation}. This progress has established AI as a central component of modern biomedical research and clinical decision support.

Yet this success conceals a fundamental limitation. Contemporary AI systems in biomedicine are trained on historical data and learn statistical associations within previously observed biological and clinical states. Their outputs---risk scores, classifications, or predicted outcomes---reflect patterns extracted from the past. As a result, these systems are inherently constrained by the distributions on which they are trained. Even as data scale increases and models become more expressive, their core mode of operation remains observational.

Medicine, however, is not an observational discipline. It is fundamentally interventional. Clinical and therapeutic decisions require reasoning about actions that alter the state of a biological system: introducing a new therapy, combining treatments, or targeting mechanisms that may not yet be fully represented in existing data. In such settings, the central question is not what is likely to happen based on past observations, but what will happen under a specific intervention. This creates a structural gap between prediction and decision-making.

As a result, the limiting factor in biomedical AI is no longer predictive accuracy, but the ability to support decision-making under intervention. This shifts the focus of AI systems from prediction to the decision layer, where therapeutic choices are made. We define the decision layer as the computational level at which alternative interventions are evaluated and selected based on modeled system behavior.

This gap cannot be closed by scale alone. Increasing the volume or diversity of data improves predictive performance within known distributions, but does not yield a representation of how biological systems evolve under perturbation. As a result, even highly accurate predictive systems fail when decision-making depends on interventions that extend beyond previously observed conditions. In systems where decisions depend on intervention rather than extrapolation, predictive architectures are structurally insufficient.

In this paper, we argue that AI in biomedicine is undergoing a structural transition. The field is moving from predictive systems toward models that explicitly represent disease as a dynamic biological process. These models capture not only associations, but the state and evolution of disease, enabling reasoning about intervention rather than extrapolation from past observations. We refer to this transition as the shift from Observational AI to Interventional AI. This shift is not a technological choice, but a structural consequence of how decisions are made in medicine.

This shift is becoming feasible due to the convergence of several factors: the availability of large-scale multimodal datasets, advances in generative modeling, and increasing computational capacity. Together, these developments enable the construction of models that go beyond static prediction toward representing and simulating biological systems.

\section{The Current Paradigm: Predictive AI}
\label{sec:predictive}

The recent progress of artificial intelligence in biomedicine has been driven by the ability to integrate and learn from large-scale, multimodal datasets. These datasets combine diverse sources of patient information, including genomic and transcriptomic profiles, histopathology and radiology images, clinical records, and treatment histories. Advances in machine learning---particularly deep learning and foundation models---have enabled the extraction of high-dimensional representations from these data, supporting a wide range of predictive tasks\citep{esteva2019guide,topol2019high,moor2023foundation}.

Within this paradigm, AI systems are designed to map complex biological and clinical inputs to specific outcomes. Typical applications include predicting disease progression, estimating survival, stratifying patients into risk groups, identifying biomarkers, and forecasting treatment response. Foundation models trained on large patient cohorts provide generalized embeddings that can be adapted to downstream tasks, further improving predictive performance across settings.

Despite their technical sophistication, these systems share a common structure. They operate by learning statistical associations between observed inputs and observed outcomes. Training is performed on retrospective datasets, where both features and labels are derived from historical observations. Model optimization is therefore guided by the ability to reproduce patterns present in past data, with performance measured through predictive accuracy on held-out samples drawn from similar distributions.

This approach has proven highly effective within its domain of applicability. When the target task is well represented in the training data, and when future cases resemble past observations, predictive models can achieve strong performance and provide meaningful clinical insights. As a result, predictive AI has become an increasingly important tool in biomedical research and clinical decision support\citep{esteva2019guide,topol2019high}.

However, the capabilities of these systems are fundamentally shaped by their observational nature. The models do not encode an explicit representation of biological processes or mechanisms; instead, they capture correlations across high-dimensional feature spaces. Their internal representations---embeddings---summarize patterns in the data but do not, in general, correspond to interpretable or mechanistic states of the underlying biological system.

As a consequence, predictive AI systems are intrinsically tied to the distributions on which they are trained. Their ability to generalize depends on the similarity between new cases and historical data, and their outputs reflect interpolation within known regimes rather than reasoning about unobserved conditions. While scaling data and model capacity improves performance, it does not change this underlying mode of operation.

In this sense, the current paradigm of AI in biomedicine can be understood as one of observational intelligence: systems that learn from what has been seen and estimate what is likely to occur under similar conditions. This paradigm has enabled substantial progress, but it also defines the boundary of what predictive AI can achieve.

\section{Structural Limits of Predictive Systems}
\label{sec:limits}

The predictive paradigm in biomedical AI has achieved substantial success, yet its limitations are not merely practical---they are structural. These systems are designed to learn mappings from observed inputs to observed outcomes within historical datasets. As such, their capabilities are inherently bounded by the distributions of data on which they are trained. This limitation does not arise from insufficient scale or model capacity, but from the underlying formulation: systems that learn from past observations cannot, by construction, represent the effects of interventions that have not yet been observed.

A central consequence of this formulation is dependence on historical distributions. Predictive models approximate conditional relationships present in past observations and generalize through interpolation within this space. When future cases resemble past data, performance can remain strong. However, when conditions shift---through new therapies, novel combinations, or emerging biological mechanisms---this mode of generalization fails, consistent with broader observations of performance degradation under distribution shift\citep{recht2019imagenet,ovadia2019trust,subbaswamy2020development}. This failure is not accidental; it is structurally inevitable in systems that do not model how interventions alter system state. Increasing data scale improves interpolation within known regimes, but does not enable reasoning about unobserved interventions. This limitation is not overcome by scaling.

For example, a predictive model trained on historical data may accurately estimate response to a PD-1 inhibitor within previously observed patient populations. However, when a novel combination therapy is introduced, the model lacks a representation of how tumor and immune states interact under the new perturbation, and its predictions degrade. A disease-level model, by contrast, can simulate the joint dynamics of tumor and immune response under combined intervention, enabling reasoning about therapeutic effects that are not directly observed in training data.

More fundamentally, predictive models do not encode the effects of intervention. While they may capture associations between treatments and outcomes, they do not represent how a biological system transitions from one state to another under perturbation. Clinical decision-making, however, depends precisely on this capability. Without an explicit representation of state, dynamics, and intervention, predictive systems cannot, by construction, answer counterfactual questions about the consequences of action.

These limitations define a structural boundary. Training on historical data constrains models to approximate previously observed outcomes, while biomedical decision-making increasingly requires reasoning beyond those observations. As the space of possible interventions expands, predictive performance within known regimes no longer guarantees decision quality. Increasing data scale or model complexity improves interpolation, but does not overcome the absence of a representation of intervention. Predictive AI can optimize within the past, but cannot, by construction, support reasoning about the future of disease under intervention, reflecting a broader distinction between statistical prediction and causal or interventional reasoning\citep{pearl2009causality,pearl2018why}.

\begin{proposition}[Structural limitation of predictive systems]
Systems trained to predict outcomes from historical data can achieve high accuracy within observed regimes, but cannot, by construction, reliably support decision-making under novel intervention.
\end{proposition}

\section{The Emerging Paradigm: Disease-Level Models}
\label{sec:emerging}

The limitations of predictive systems point toward an emerging paradigm in biomedical AI: models that explicitly represent disease as a dynamic biological process. Rather than learning associations between inputs and outcomes, these systems aim to capture the state, structure, and evolution of disease across molecular, cellular, and clinical dimensions. This shift reflects a move from correlation-based inference to models capable of supporting reasoning about intervention.

At the core of this paradigm is the concept of a biological state space. In this view, disease is not defined by static features or isolated measurements, but by the joint configuration of interacting components---such as tumor cells, immune populations, signaling pathways, and the surrounding microenvironment. A disease state corresponds to a position within this high-dimensional space, while disease progression reflects trajectories through it over time.

Disease-level models are designed to operate over this state space. Rather than mapping inputs directly to outcomes, they seek to represent how the system evolves and responds to perturbations. This requires moving beyond feature aggregation toward explicit modeling of interactions and dynamics. In such systems, predictions are not simply outputs of a learned function, but consequences of modeled changes in system state.

\begin{figure}[htbp]
  \centering
  \includegraphics[width=\textwidth]{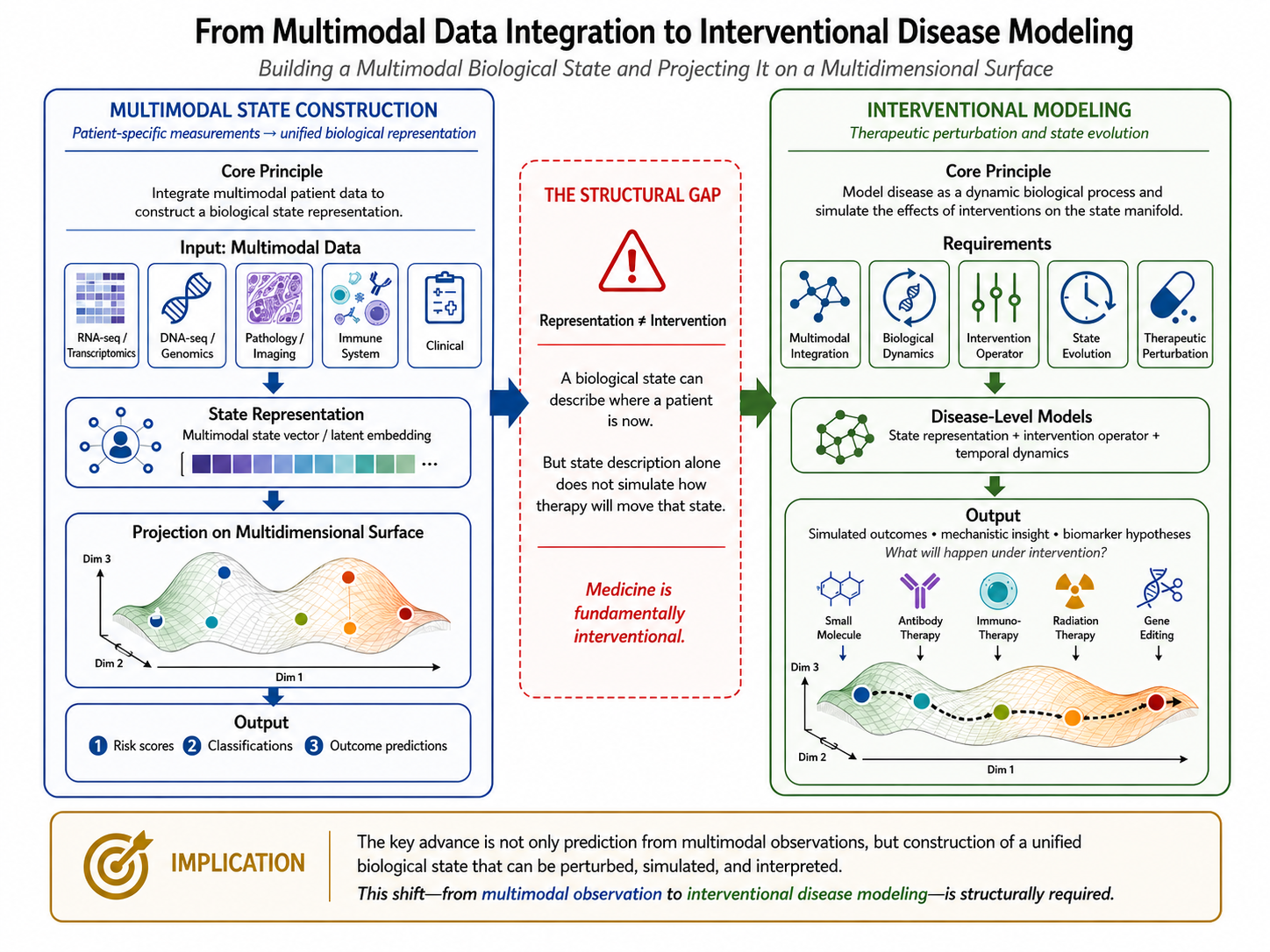}
  \caption{\textbf{From observational to interventional intelligence in biomedicine.}
  Current AI systems in biomedicine operate within an observational paradigm, learning statistical associations from historical multimodal data to generate predictions such as risk scores, classifications, and patient stratification. While highly effective within previously observed distributions, these systems are inherently constrained in their ability to reason about novel interventions. The emerging paradigm shifts toward interventional intelligence, in which models explicitly represent disease as a dynamic biological process and simulate its evolution under therapeutic perturbations. This transition---from prediction to simulation, and from correlation to modeled dynamics---enables reasoning about counterfactual scenarios and supports decision-making under conditions not directly observed in training data.}
  \label{fig:paradigm}
\end{figure}

In this sense, disease-level models define a shift from observational intelligence to interventional intelligence. They provide a framework in which AI systems can move beyond describing what has happened to reasoning about what can happen, and under which conditions. This structural transition---from Observational AI to Interventional AI---is illustrated schematically in Figure~\ref{fig:paradigm}, which contrasts correlation-based inference with simulation-based reasoning.

\begin{figure}[htbp]
  \centering
  \includegraphics[width=\textwidth]{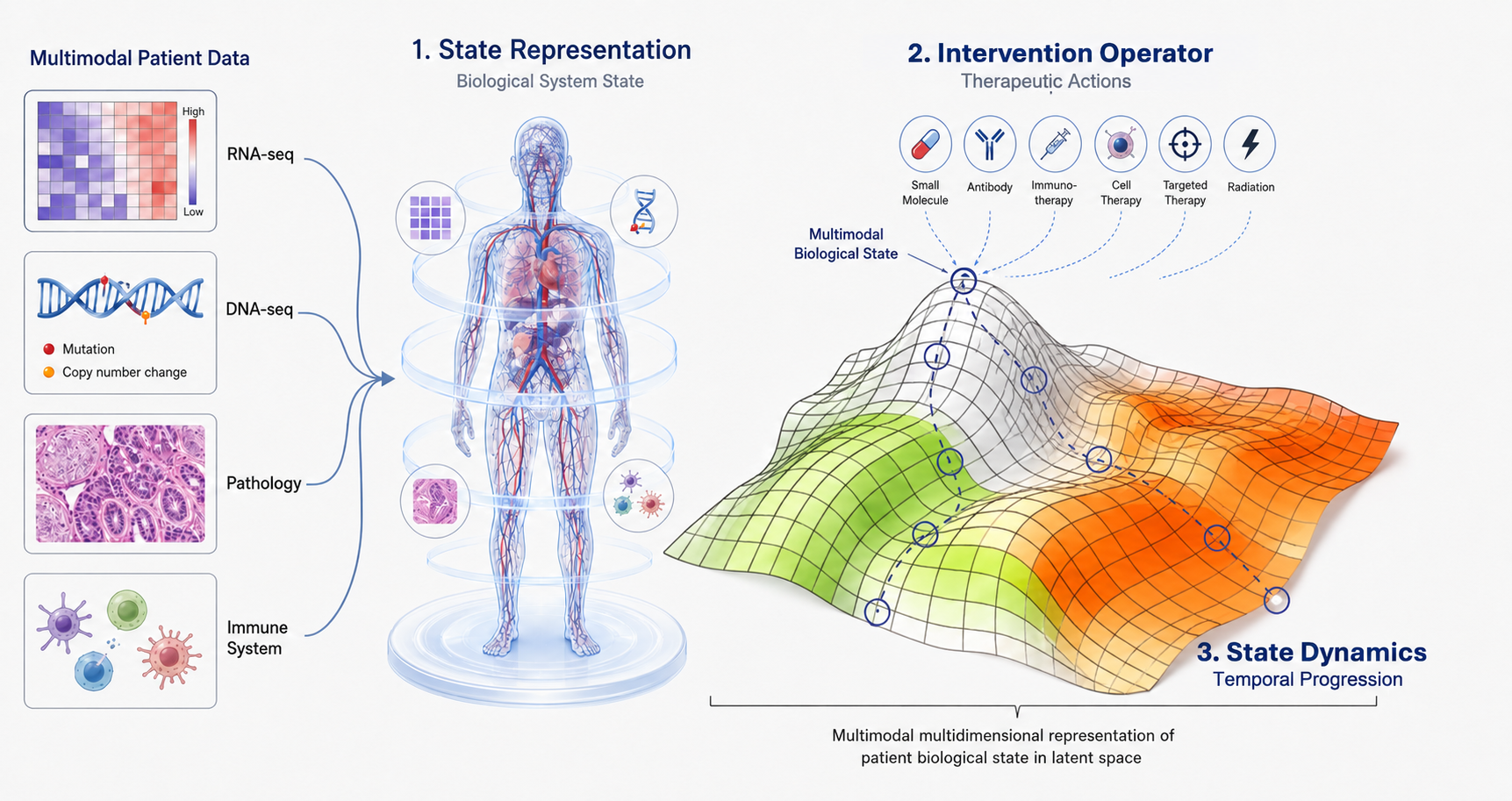}
  \caption{\textbf{The Disease Model Triad.}
  Disease-level models are defined by three interdependent components: state representation, which encodes the biological configuration of the system across tumor, immune, and microenvironmental dimensions; state dynamics, which describe how this state evolves over time under endogenous processes; and the intervention operator, which maps therapeutic actions to transitions in system state. Together, these components define a minimal computational architecture for interventional intelligence, enabling models to simulate disease trajectories under perturbation and to reason about counterfactual outcomes beyond observed data.}
  \label{fig:triad}
\end{figure}

To support this transition computationally, disease-level models must satisfy a more specific structural requirement. We refer to this structure as the Disease Model Triad (Figure~\ref{fig:triad}), consisting of three interdependent components:

\begin{itemize}[leftmargin=*]
  \item \textbf{State Representation}, which encodes the joint configuration of tumor, immune system, and microenvironment as a coherent biological state within a high-dimensional but structured space.
  \item \textbf{State Dynamics}, which describe how this state evolves over time under endogenous processes and external influences, capturing trajectories of disease progression.
  \item \textbf{Intervention Operator}, which maps therapeutic actions to transitions in system state, enabling the modeling of how interventions perturb and redirect disease trajectories.
\end{itemize}

The Disease Model Triad defines the minimal structure required for any AI system to support reasoning under intervention.

Together, these components define a system capable of reasoning about intervention by simulating state transitions under perturbation. Rather than estimating outcomes from historical correlations, the model represents a current state, applies a perturbation, and computes the resulting evolution of the system. Predictions emerge as consequences of simulated state transitions rather than direct outputs of statistical association.

This triadic structure defines the minimal computational architecture required for interventional intelligence in biomedicine. Models that lack any of these components may achieve strong predictive performance, but remain structurally limited in their ability to support reasoning about intervention.

This paradigm introduces a different objective for model development. Instead of optimizing solely for predictive accuracy, disease-level models are evaluated by their ability to faithfully represent biological structure and dynamics, and to support reasoning about interventions that may not be directly observed in training data. As a result, model performance becomes tied not only to statistical fit, but to the quality of the underlying representation of disease.

Importantly, this transition does not eliminate the role of data-driven learning. Multimodal data remain essential for constructing and constraining these models. However, data are no longer the endpoint; they become the substrate for learning representations of biological systems that can be interrogated, perturbed, and simulated. This structure does not merely extend predictive systems---it defines a different computational regime.

\section{From Prediction to Simulation}
\label{sec:simulation}

The shift from prediction to simulation is not a change in output, but a change in the computational primitive underlying biomedical reasoning. Predictive AI answers questions of the form ``what is likely to happen?'' based on patterns observed in historical data. Disease-level models, by contrast, address a different class of questions: ``what will happen if a specific intervention is applied?'' This distinction marks a fundamental change in the role of AI in biomedicine (Table~\ref{tab:comparison}).

\begin{table}[htbp]
  \centering
  \caption{Observational AI vs Interventional AI}
  \label{tab:comparison}
  \begin{tabular}{lll}
    \toprule
    \textbf{Aspect} & \textbf{Observational} & \textbf{Interventional} \\
    \midrule
    Core question   & What is likely?        & What will happen if?    \\
    Data basis      & Historical             & Modeled + data          \\
    Output          & Prediction             & Simulation              \\
    Limitation      & Distribution-bound     & Model-bound             \\
    \bottomrule
  \end{tabular}
\end{table}

Simulation becomes the central computational primitive of biomedical AI when decisions depend on intervention. In this sense, simulation is not a feature of the system, but its defining computational operation: just as prediction defines observational AI, simulation defines interventional AI.

\begin{figure}[htbp]
  \centering
  \includegraphics[width=\textwidth]{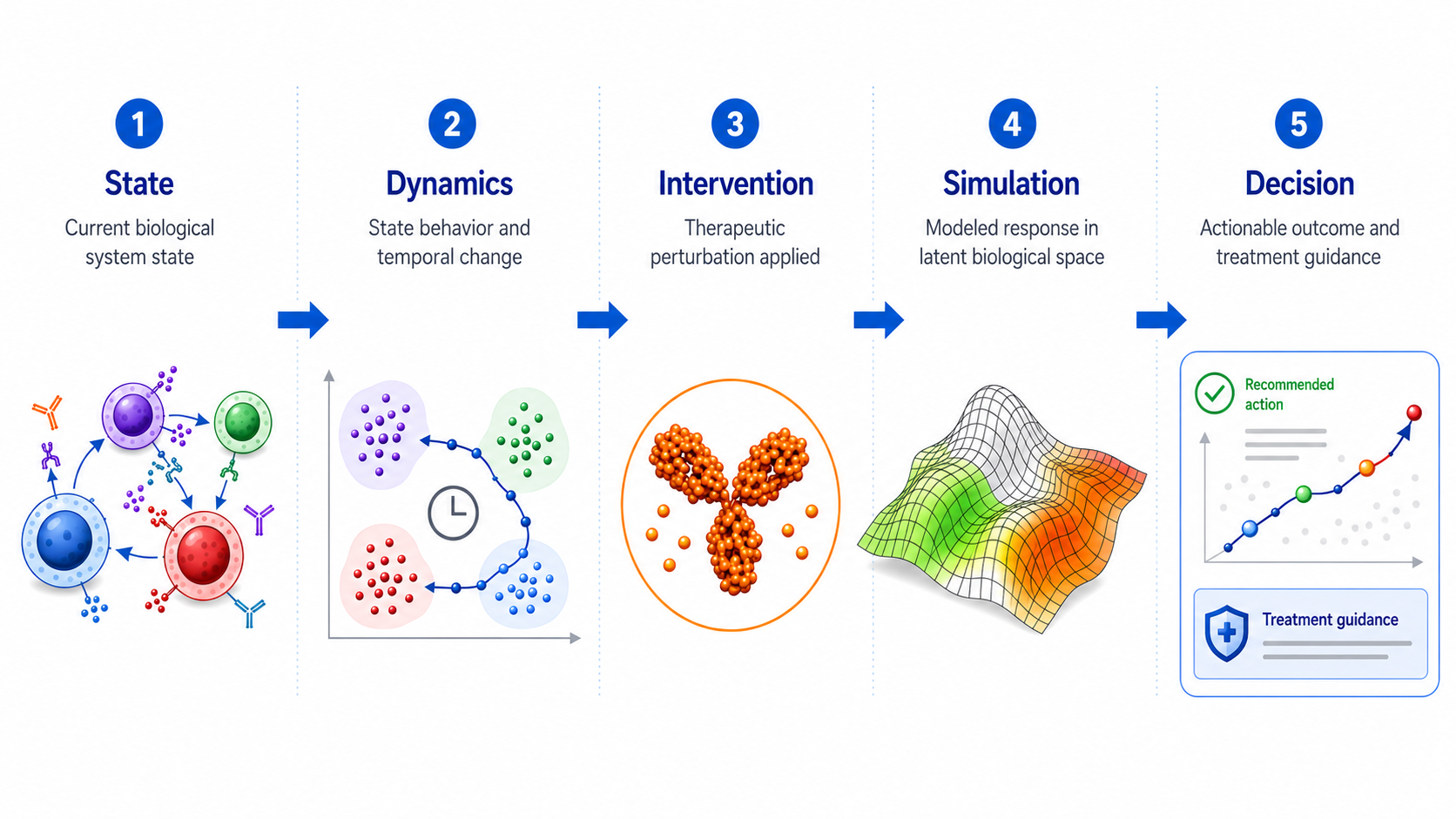}
  \caption{\textbf{From state representation to decision through simulation.}
  Disease-level models operate by representing the biological state of a system, modeling its dynamics, and applying interventions as perturbations to that state. Simulation of these perturbations generates trajectories that reflect how disease evolves under different therapeutic actions. Decisions are then derived from these simulated trajectories rather than inferred from historical correlations. This framework captures the transition from observational to interventional intelligence, where reasoning about intervention replaces prediction based on past observations.}
  \label{fig:pipeline}
\end{figure}

Rather than mapping inputs directly to predictions, models operate by representing a system state, applying a perturbation, and computing the resulting transition (Figure~\ref{fig:pipeline}). In this framework, predictions are no longer direct outputs of statistical association, but emerge as consequences of modeled system dynamics. This formulation enables reasoning about how disease evolves under intervention, rather than extrapolation from past observations.

As Figure~\ref{fig:pipeline} illustrates, decision-making in this framework is a consequence of simulated system evolution rather than direct inference from data. This shift introduces a counterfactual mode of reasoning that is largely absent from predictive systems, aligning with formal frameworks of causal inference and intervention-based reasoning\citep{pearl2009causality,pearl2018why}. In clinical and research settings, decision-making depends on evaluating scenarios that have not yet been observed: introducing a new therapy, combining treatments, altering dosage, or targeting a specific biological pathway. These scenarios cannot be reliably addressed through interpolation within historical data alone. Simulation provides a mechanism for exploring such possibilities by enabling models to reason over unobserved interventions, an approach increasingly explored in generative and world-model-based systems\citep{lecun2022path,ha2018world}.

Importantly, simulation does not imply exact mechanistic modeling in the traditional sense. Instead, it reflects the ability of a system to generate consistent and biologically plausible trajectories under perturbation, informed by both data and learned representations of structure and dynamics, as seen in emerging generative modeling approaches\citep{lecun2022path,ha2018world}. The goal is not to replace empirical observation, but to augment it with a framework for systematic exploration of the intervention space.

As a result, the function of AI shifts from predictive estimation to decision support grounded in modeled behavior. Outputs are no longer limited to risk scores or classifications, but include simulated disease trajectories, projected responses to therapy, and comparative evaluations of alternative interventions. This expands the role of AI from an analytical tool to a system for reasoning about action.

The implications of this transition are significant. When simulation becomes central, the value of a model is determined not only by its predictive accuracy, but by its capacity to support reliable reasoning under intervention. This redefines both the objectives of model development and the criteria by which models are evaluated.

In this emerging paradigm, prediction remains a component of the system, but no longer defines it. The primary capability is the ability to represent, perturb, and simulate biological systems in a way that supports informed decision-making in the presence of uncertainty and novelty.

This distinction is not incremental---it defines the boundary between systems that predict within the past and systems that reason about intervention.

\section{Implications for Drug Development}
\label{sec:drug}

The transition from predictive AI to disease-level models has significant implications for drug development, where decision-making depends on navigating a vast and largely unobserved space of potential interventions. Traditional approaches rely heavily on empirical iteration: hypotheses are tested sequentially through preclinical studies and clinical trials, with outcomes informing subsequent decisions. Predictive AI has improved this process by enabling better patient stratification, biomarker discovery, and outcome forecasting. However, these improvements remain anchored in historical data and observed treatment effects.

Disease-level models introduce a different mode of operation. By representing the state and dynamics of biological systems, these models enable reasoning about interventions that have not yet been tested empirically. This capability is particularly relevant in areas where combinatorial complexity is high, such as oncology, where multiple pathways interact and therapeutic strategies often involve combinations of agents. The number of possible interventions grows rapidly beyond what can be explored through trial-and-error approaches alone.

In this context, simulation-informed drug development becomes structurally necessary. Models can be used to explore the intervention space by generating and evaluating hypotheses in silico before committing to experimental or clinical validation. Potential therapies, combinations, and dosing strategies can be assessed based on their projected impact on disease dynamics, narrowing the search space and prioritizing the most promising candidates.

This shift has implications across multiple stages of the development pipeline. In early discovery, disease-level models can support target identification by revealing system-level vulnerabilities and interaction patterns. In preclinical development, they can guide the design of experiments by identifying conditions under which specific mechanisms are likely to be effective. In clinical development, they can inform trial design by predicting how different patient subpopulations will respond to interventions, enabling more precise inclusion criteria and adaptive trial strategies.

Patient selection, in particular, moves from a correlational to a mechanistic basis. Instead of identifying cohorts based solely on observed associations, models can stratify patients according to underlying biological states and predicted response to intervention. This allows for more targeted trials and increases the likelihood of detecting treatment effects.

More broadly, drug development shifts from empirical iteration over candidate therapies to model-guided exploration of the intervention space. This effectively moves decision-making upstream, from late-stage clinical trials toward earlier, simulation-informed evaluation. While experimental validation remains essential, the role of AI evolves from analyzing outcomes to shaping the sequence of decisions that lead to them. As a result, the efficiency and direction of the development process become increasingly dependent on the quality of the underlying disease model.

This transition does not eliminate uncertainty, but it changes how uncertainty is managed. Rather than relying solely on retrospective evidence, developers can use simulation to evaluate competing hypotheses, assess risks, and design experiments that are more informative. In doing so, disease-level models redefine the interface between data, experimentation, and decision-making in drug development.

\section{Implications for Clinical Care}
\label{sec:clinical}

The transition from predictive systems to disease-level models has equally significant implications for clinical care, where decisions must be made at the level of individual patients under conditions of uncertainty. Current applications of AI in the clinic are largely predictive: they estimate risk, classify disease subtypes, or forecast outcomes based on patterns observed in historical data. While these tools can support clinical judgment, they do not fundamentally alter its structure. Decisions are still made by combining retrospective evidence, clinical guidelines, and physician experience.

This paradigm reflects the observational nature of predictive AI. Models provide estimates of what is likely to occur under conditions similar to those seen before, but they do not directly address the central question of clinical practice: what should be done for this patient, under these specific circumstances, given the available interventions. As a result, predictive systems often function as decision-support tools rather than decision-reasoning systems.

Disease-level models enable a different mode of interaction. By representing the biological state of an individual patient and modeling how that state evolves under intervention, these systems can support reasoning about treatment choices. Instead of relying solely on population-level correlations, clinicians can evaluate how alternative therapies are expected to influence the trajectory of disease in a specific patient context.

This shift has several implications for clinical practice. First, decision support moves from static to dynamic. Rather than providing fixed risk scores or classifications, models can generate evolving assessments as new data become available and as the patient's condition changes over time. Second, personalization becomes mechanistic rather than correlational. Treatment recommendations are informed by the modeled biological state and its response to intervention, rather than by similarity to historical cohorts alone.

Third, clinical reasoning becomes increasingly prospective. Instead of relying primarily on retrospective evidence and guidelines derived from aggregate populations, clinicians can use simulation to evaluate potential treatment strategies before they are applied. This allows for more informed exploration of alternatives, particularly in complex cases where standard protocols may not fully capture the relevant biological variability.

Importantly, this transition does not replace clinical expertise. Rather, it augments it by providing a structured framework for reasoning about intervention. Physicians remain responsible for interpreting model outputs, integrating them with clinical context, and making final decisions. However, the nature of decision-making evolves: from selecting among predefined options based on past evidence, to evaluating modeled outcomes under different intervention scenarios.

More broadly, clinical care shifts from decision support to decision reasoning. AI systems move from providing estimates about the future to enabling structured exploration of possible futures under intervention. In doing so, they redefine the role of AI in medicine---from tools that assist in interpretation to systems that actively participate in the reasoning process underlying clinical decisions.

\section{Architectural Requirements}
\label{sec:architecture}

The transition from predictive systems to disease-level models cannot be achieved by scaling existing architectures. It requires a fundamentally different system design. It defines a fundamentally different class of system. While advances in data integration and model capacity have driven the success of predictive AI, disease-level modeling requires systems that explicitly represent biological structure, dynamics, and response to intervention. This shift introduces new constraints on how models are designed, trained, and evaluated.

First, multimodal integration remains necessary but is no longer sufficient. Predictive systems often treat multimodal data as complementary feature sets to be aggregated into a shared representation. In contrast, disease-level models require integration that preserves the relationships between modalities, reflecting how genomic, transcriptomic, cellular, and clinical factors interact within a unified biological system. The objective is not simply to combine data, but to represent the dependencies and interactions that define disease behavior.

Second, models must capture biological dynamics rather than static associations. Disease progression unfolds over time through evolving interactions among components of the system. Architectural approaches must therefore support longitudinal modeling, enabling the representation of state transitions and temporal dependencies. This includes the ability to incorporate sequential data, track trajectories through biological state space, and model how systems evolve under both endogenous processes and external perturbations.

Third, the capacity to model intervention is essential. Disease-level models must represent how therapeutic actions alter system state, requiring explicit encoding of perturbations and their effects. This extends beyond associating treatments with outcomes to modeling the mechanisms by which interventions influence biological processes. Architectures must therefore support counterfactual reasoning, enabling the evaluation of alternative intervention strategies that may not be directly observed in training data.

Fourth, generative modeling becomes central. Approaches such as diffusion models, latent-variable models, and other generative frameworks provide a means to represent distributions over system states and simulate transitions under perturbation. These methods enable the construction of models that can generate plausible trajectories of disease evolution, rather than merely predicting endpoints. The emphasis shifts from discriminative performance to the fidelity of generated dynamics.

Fifth, representations must be structured and interpretable at the level of biological systems. While high-dimensional embeddings are effective for prediction, disease-level models require representations that correspond to meaningful biological entities and processes. This does not imply full interpretability in a traditional sense, but it does require alignment between model components and the underlying structure of the system being modeled, enabling validation, interrogation, and refinement.

Finally, these requirements introduce new criteria for model evaluation. Performance can no longer be assessed solely through predictive accuracy on held-out data. Instead, evaluation must consider the consistency of simulated trajectories, the plausibility of intervention responses, and the model's ability to generalize to unobserved scenarios. This may involve integrating experimental validation, prospective studies, and domain-specific benchmarks that reflect the goals of simulation and intervention.

Together, these components enable reasoning about intervention through simulation of state transitions. As such, their development requires rethinking both the objectives and the structure of AI systems in biomedicine.

These architectural requirements are not optional extensions of predictive systems, but necessary conditions for supporting reasoning under intervention.

\section{Open Challenges}
\label{sec:challenges}

The transition from predictive systems to disease-level models introduces a set of challenges that are both technical and conceptual. While the potential of this paradigm is significant, its realization requires addressing limitations that are not fully resolved by current approaches. These challenges define an active area of research at the intersection of machine learning, biology, and clinical practice.

A central challenge is the validation of generative biological models. Unlike predictive systems, which can be evaluated through standard metrics such as accuracy or area under the curve, disease-level models must be assessed based on their ability to represent dynamics and simulate intervention. This requires new evaluation frameworks that go beyond retrospective validation. Experimental studies, prospective trials, and targeted perturbation experiments may be necessary to determine whether model-generated trajectories correspond to real biological behavior. Establishing reliable validation protocols remains an open problem.

Data limitations represent a second major constraint. While large-scale multimodal datasets have enabled advances in predictive AI, data capturing interventional effects---particularly at the level of molecular and cellular dynamics---are relatively sparse. Clinical data are often observational, heterogeneous, and incomplete with respect to the variables required for modeling system dynamics. Preclinical and experimental datasets provide richer information about mechanisms but may not fully translate to human biology. Bridging these data sources, and learning from them in a coherent framework, remains a key challenge.

A related issue is the integration of models with clinical workflows. For disease-level models to be useful in practice, they must operate within the constraints of healthcare systems, including data availability, time pressures, regulatory requirements, and clinician usability. Outputs must be interpretable and actionable, and systems must support decision-making without introducing excessive complexity or uncertainty. Aligning advanced modeling capabilities with real-world clinical environments is a nontrivial task.

The trade-off between interpretability and fidelity also becomes more pronounced. Models that capture complex biological dynamics may rely on high-dimensional latent representations that are difficult to interpret directly. At the same time, clinical adoption often depends on transparency and trust. Balancing the need for expressive, high-fidelity models with the requirement for understandable and verifiable outputs is an ongoing challenge.

Finally, there are broader questions related to generalization and robustness. Disease-level models aim to operate in regimes that extend beyond observed data, making them inherently sensitive to assumptions about system structure and dynamics. Ensuring that these models remain reliable under novel conditions, and that their predictions do not degrade in unexpected ways, requires careful design, validation, and monitoring.

These challenges do not diminish the importance of the emerging paradigm; rather, they define the work required to realize it. Addressing them will require advances not only in machine learning, but also in experimental biology, data generation, and the integration of computational systems into clinical and research practice.

\section{Conclusion}
\label{sec:conclusion}

Artificial intelligence has already transformed biomedicine by enabling the extraction of signal from complex, multimodal data and supporting increasingly accurate prediction of clinical outcomes. This progress has established predictive AI as a powerful analytical tool. However, as biomedical decision-making increasingly depends on reasoning about intervention rather than extrapolation from past observations, the limitations of the predictive paradigm become structural rather than incremental.

The field is therefore undergoing a fundamental transition---from Observational AI to Interventional AI. The next generation of AI systems in biomedicine will not be defined by improved prediction alone, but by their capacity to represent disease as a dynamic biological process, to model its evolution, and to evaluate the effects of intervention. This shift from observational to interventional intelligence follows directly from the structure of biomedical decision-making and is not a matter of technological preference.

This distinction defines a boundary. Systems that learn from historical data remain effective within known regimes, but become increasingly unreliable as the space of possible interventions expands. Systems that explicitly represent state, dynamics, and intervention provide a fundamentally different capability: the ability to reason about possible futures rather than infer from past observations.

Increasing data scale and model complexity will continue to improve predictive performance, but will not overcome the absence of a representation of intervention. Systems that learn from the past remain confined to it. Systems that model intervention define the future of medicine.

This transition also has implications for how value is created and captured in biomedical AI. Systems that support decision-making under intervention operate at a different level than predictive tools: they do not merely inform decisions, but increasingly define how decisions are made. As such systems become embedded in drug development and clinical workflows, they are likely to become central to the economics of biomedical innovation.

The question is no longer whether AI will influence decisions in medicine, but which systems will define them---and therefore which systems will control the decision layer.

The future of AI in biomedicine will be defined by the ability to represent, simulate, and ultimately control biological systems under intervention. This transition is not optional: systems that cannot model intervention will be structurally excluded from decision-making.



{\footnotesize
\begin{thebibliography}{10}
\providecommand{\natexlab}[1]{#1}
\providecommand{\url}[1]{\texttt{#1}}
\expandafter\ifx\csname urlstyle\endcsname\relax
  \providecommand{\doi}[1]{doi: #1}\else
  \providecommand{\doi}{doi: \begingroup \urlstyle{rm}\Url}\fi

\bibitem[Esteva et~al.(2019)Esteva, Robicquet, Ramsundar, Kuleshov, DeFauw,
  Chou, Cui, Corrado, Thrun, and Dean]{esteva2019guide}
Andre Esteva, Alexandre Robicquet, Bharath Ramsundar, Volodymyr Kuleshov,
  Jeffrey DeFauw, Katherine Chou, Claire Cui, Greg Corrado, Sebastian Thrun,
  and Jeff Dean.
\newblock A guide to deep learning in healthcare.
\newblock \emph{Nature Medicine}, 25:\penalty0 24--29, 2019.
\newblock \doi{10.1038/s41591-018-0316-z}.

\bibitem[Ha and Schmidhuber(2018)]{ha2018world}
David Ha and J{\"u}rgen Schmidhuber.
\newblock World models.
\newblock In \emph{Advances in Neural Information Processing Systems
  (NeurIPS)}, 2018.

\bibitem[LeCun(2022)]{lecun2022path}
Yann LeCun.
\newblock A path towards autonomous machine intelligence.
\newblock Technical report, 2022.
\newblock URL \url{https://openreview.net/pdf?id=BZ5a1r-kVsf}.

\bibitem[Moor et~al.(2023)Moor, Banerjee, Abad, Krber, Lozano, Langlotz,
  et~al.]{moor2023foundation}
Michael Moor, Oishi Banerjee, Zahra Shakeri~Hossein Abad, Harlan~M. Krber,
  Aurelio Lozano, Curtis~P. Langlotz, et~al.
\newblock Foundation models for generalist medical artificial intelligence.
\newblock \emph{Nature}, 616:\penalty0 259--265, 2023.
\newblock \doi{10.1038/s41586-023-05881-4}.

\bibitem[Ovadia et~al.(2019)Ovadia, Fertig, Ren, Nado, Sculley, Nowozin,
  Dillon, Lakshminarayanan, and Snoek]{ovadia2019trust}
Yaniv Ovadia, Emily Fertig, Jie Ren, Zachary Nado, D.~Sculley, Sebastian
  Nowozin, Joshua~V. Dillon, Balaji Lakshminarayanan, and Jasper Snoek.
\newblock Can you trust your model's uncertainty? {E}valuating predictive
  uncertainty under dataset shift.
\newblock In \emph{Advances in Neural Information Processing Systems
  (NeurIPS)}, 2019.

\bibitem[Pearl(2009)]{pearl2009causality}
Judea Pearl.
\newblock \emph{Causality: Models, Reasoning, and Inference}.
\newblock Cambridge University Press, 2nd edition, 2009.

\bibitem[Pearl and Mackenzie(2018)]{pearl2018why}
Judea Pearl and Dana Mackenzie.
\newblock \emph{The Book of Why: The New Science of Cause and Effect}.
\newblock Basic Books, 2018.

\bibitem[Recht et~al.(2019)Recht, Roelofs, Schmidt, and
  Shankar]{recht2019imagenet}
Benjamin Recht, Rebecca Roelofs, Ludwig Schmidt, and Vaishaal Shankar.
\newblock Do {ImageNet} classifiers generalize to {ImageNet}?
\newblock In \emph{International Conference on Machine Learning (ICML)}, 2019.

\bibitem[Subbaswamy and Saria(2020)]{subbaswamy2020development}
Adarsh Subbaswamy and Suchi Saria.
\newblock From development to deployment: Dataset shift, causality, and
  shift-stable models in health {AI}.
\newblock \emph{Biostatistics}, 21\penalty0 (2):\penalty0 345--352, 2020.
\newblock \doi{10.1093/biostatistics/kxz041}.

\bibitem[Topol(2019)]{topol2019high}
Eric~J. Topol.
\newblock High-performance medicine: The convergence of human and artificial
  intelligence.
\newblock \emph{Nature Medicine}, 25:\penalty0 44--56, 2019.
\newblock \doi{10.1038/s41591-018-0300-7}.

\end{thebibliography}

}
\end{document}